\documentclass[reprint,prd,nofootinbib]{revtex4-2}

\usepackage{nameref}

\usepackage{fancyhdr}
\newcommand{\changefont}{%
	\fontsize{8}{11}\selectfont
}

\usepackage{soul}
\usepackage[dvipsnames]{xcolor}
\usepackage{amsmath}
\usepackage{amsfonts}
\usepackage{float}
\usepackage{subcaption}
\usepackage{multirow}
\usepackage{hyperref}
\hypersetup{
    colorlinks=true,
    linkcolor=black,
    citecolor=CadetBlue,
    filecolor=CadetBlue,      
    urlcolor=CadetBlue,
}
\usepackage[skip=10pt plus1pt]{parskip}
\usepackage{lipsum}
\usepackage{tikz}
\usepackage{ragged2e}
\usepackage{animate}
\usepackage{graphicx}
\usepackage{array}
\usepackage{hhline}
\usepackage{colortbl}
\usepackage{tabularx}
\usepackage{pgfplots}

\usepackage{braket}
\usepackage{xspace}
\usepackage{tikz}
\usepackage{tikz-cd}

\begin{document}
\title{Large Spin Systematics: Patterns from Reciprocity for Multiple Spinning Operators}

\author{Pulkit Agarwal}
\affiliation{Centro de F\'{i}sica do Porto, Faculdade de Ci\^{e}ncias da Universidade do Porto, Portugal}
\date{\today}
\begin{abstract}
    We study the behaviour of the conformal block expansions of scalar fivepoint Lorentzian conformal correlators in the limit where multiple cross ratios approach zero. Since this limit is controlled by intermediate operators with large spin, we use it to study the large spin expansion of the OPE coefficients involving these operators. By imposing bootstrap assumptions such as analyticity of the correlators, we derive an infinite set of new constraints on the large spin behaviour of OPE coefficients involving multiple spinning operators. We also show that for the case of $l=0$, these constraints can be trivialised to all orders in $1/J$ by identifying a pattern in the coefficients.
\end{abstract}
\maketitle

\section{Introduction}
In recent years, significant progress has been made in our understanding of Conformal Field Theories (CFTs). We know that correlators in any given CFT are completely characterised by the scaling dimensions of the operators present in the theory, along with their three point functions. From this data, one can reconstruct any higher point correlator using the Operator Product Expansion (OPE). For example, one can write the fourpoint correlator as a product of two three point functions with a sum over an infinite set of intermediate two point functions. Of course, due to conformal symmetry, operators organise themselves into families of primaries and their descendants, and the sum over intermediate states can be rewritten as a sum over ``conformal blocks'' associated with each primary $\mathcal{O}$: 
\begin{equation}\label{eq:4ptfn}
    \frac{\braket{\mathcal{O}_1\mathcal{O}_2\mathcal{O}_3\mathcal{O}_4}}{\braket{\mathcal{O}_1\mathcal{O}_2}\braket{\mathcal{O}_3\mathcal{O}_4}}\equiv \mathcal{G}(u,v) =\sum_{\Delta,\ell} a_{\Delta,\ell} G_{\Delta,\ell}(u,v).
\end{equation}
Here, $\mathcal{O}_i$ is an operator located at $x_i$, and we have redefined the product of OPE coefficients $f_{12\mathcal{O}} f_{34\mathcal{O}} = a_{\Delta,\ell}$. This makes explicit that the sum is over the quantum numbers of the intermediate operator $\mathcal{O} $. The variables $u,v$ are cross ratios defined by $u = \frac{x^2_{12}x^2_{34}}{x^2_{13}x^2_{24}}, v = \frac{x^2_{14}x^2_{23}}{x^2_{13}x^2_{24}}$. 

While the blocks are completely fixed by conformal symmetry, the \textit{partial wave amplitudes} $a_{\Delta,\ell}$ are dynamical quantities that depend on the specific CFT under consideration\footnote{We have also suppressed the identity to focus solely on exchange diagrams.}. The other piece of the CFT data is the spectrum of operators present in the theory. It is well known that this spectrum is defined by the scaling dimensions $\Delta = \Delta_0 + \gamma(\ell)$, which depends on the spin $\ell$ of the operator through the anomalous dimension $\gamma(\ell)$ \cite{Korchemsky:1988si,Belitsky:2006en,Basso:2006nk}. Here, $\Delta_0$ is the scaling dimension at the tree level\footnote{We can also define this via the conformal twist $\tau = \Delta - \ell$. The anomalous dimension is then defined by $\gamma(\ell) = \tau - \tau_0$ where $\tau_0$ is the bare twist.}.

Studying this \textit{conformal block expansion} of the scalar fourpoint function gives us access to the OPE coefficients involving two scalar external operators and one intermediate operator of arbitrary spin. Further, in the Lorentzian setting, one gains access to the limit $x^2_{12},x^2_{23} \rightarrow 0$ (or equivalently $u,v\rightarrow 0$) which is dominated by large spin. This limit has been studied extensively in the lighcone bootstrap program \cite{Alday:2007mf,Fitzpatrick:2012yx,Komargodski2013,Alday:2013cwa,Alday:2015ota,Kaviraj:2015cxa,Kaviraj:2015xsa,Simmons-Duffin:2016wlq,caron2017analyticity,simmons2018spacetime,kravchuk2018light}.

This large spin limit was also studied by \cite{Alday:2015eya} to prove the reciprocity principle which states that the large spin expansion of the anomalous dimension $\gamma(\ell)$ only contains even powers of the Casimir spin $J^2 = (\Delta + \ell)(\Delta + \ell - 2)/4$. They further constrained the large spin expansion of the partial wave amplitudes $a_{\Delta,\ell}$ (or, equivalently, the OPE coefficients) by imposing bootstrap assumptions on the correlator in this limit.

To extend these results to the OPE involving multiple spinning operators using the fourpoint function is not really efficient as, in order to introduce additional spinning operators, one typically has to use weight shifting operators \cite{Karateev:2017jgd} recursively with scalar conformal blocks as a starting point. A workaround for this problem was highlighted in \cite{Bercini:2020msp,Buric:2020dyz,Buric:2021ywo,Buric:2021ttm,Bercini:2021jti,Antunes:2021kmm,Buric:2021kgy,Bercini:2022gvs,Kaviraj:2022wbw,Poland:2023vpn,Poland:2023bny,Antunes:2023kyz,Bercini:2024pya,Bargheer:2024hfx} where it was argued that higher point functions, since they contain an infinitely many fourpoint functions, can be used to study this interesting limit.

In this paper, we will study the conformal block expansion of higher point functions of identical scalars to derive constraints in the large spin expansion of OPE coefficients involving multiple spinning operators. First, we find five point conformal blocks in the large spin expansion (Eq. \ref{eq:fiveptblock}). We next assume an ansatz for the OPE coefficients that connects smoothly with the tree-level result and, using the blocks derived earlier, we derive constraints on this ansatz order-by-order in $1/J_i$ (Eq. \ref{eq:cons1}). These constraints are valid for any tensor structure $l$ and can be computed algorithmically to any desired order in $1/J_i$. We next show that, by extracting an appropriate prefactor, it is possible to completely trivialise the constraints at $l=0$ to all orders in $1/J_i$ (Eqs. \ref{eq:cons2}, \ref{eq:cons3}). That is, all coefficients of odd powers in the large spin expansion vanish identically for $l=0$ with this new ansatz. Our results are applicable for any CFT in $d=4$ in both the perturbative and non-perturbative regimes.

\section{Kinematics}
The kinematics of $n$ point correlation functions in a CFT in $d\geq n-2$ is governed by $n(n-3)/2$ cross ratios \cite{Ginsparg:1988ui}. As illustrated in Eq. \ref{eq:4ptfn}, the \textit{reduced} fourpoint function $\mathcal{G}(u,v)$ (i.e. the appropriately renormalised correlator) only depends on these cross ratios. The conformal block expansion is then a statement of completeness of the set of conformal blocks $G_{\Delta,\ell}(u,v)$ in this space of functions of cross ratios.

In this reframing of the problem, the conformal blocks are harmonic functions over the conformal group and are therefore eigenfunctions of the Casimir operators. These functions are very well studied in the literature, both in the Euclidean and Lorentzian settings \cite{Dolan:2003hv,Dolan:2004iy,Dolan:2011dv,Hogervorst:2013sma,Costa:2011dw,Costa:2012cb,Raben:2018rbn,Agarwal:2023xwl}. In the present work, we will work exclusively in the Lorentzian setting.

Special limits of the conformal block expansion can give us insights into the spectrum of operators by narrowing down our analysis into regions dominated by interesting subclasses of the operators present. The limit $x_{12}^2\rightarrow 0$, or equivalently $u\rightarrow 0$, is dominated by the leading twist contribution and the so-called \textit{lightcone} conformal block expansion in this limit reads:
\begin{equation}
    \lim_{u\rightarrow 0}\mathcal{G}(u,v) = \sum_{\ell} a_{\ell}\ u^{\tau/2} (1-v)^{\ell}\ F_{\ell+\tau/2}(1-v)
\end{equation}
where $F_{\beta}(x) =\,_2F_1(\beta,\beta;2\beta;x)$. If we further take the limit $x_{23}^2\rightarrow 0$ (or $v\rightarrow 0$), we reach the double lightcone limit which is dominated by large spin $\ell$. In this limit, the conformal blocks simplify further into an expansion in terms of the Bessel functions \cite{Fitzpatrick:2012yx,Komargodski2013}. This is best seen by defining a rescaled conformal spin variable $j^2 = vJ^2$ and solving the lightcone Casimir equation for fixed large $\ell$ and $j/\ell\ll 1$. Further, we will remove a prefactor of the form:
\begin{equation}
    \texttt{pre}_4 = \frac{2^{-\ell}\Gamma(2\ell+\tau-1)}{\Gamma(\ell+\tau/2)^2}\ .
\end{equation}
This expansion then takes the form:
\begin{equation}\label{eq:fourptansatz}
    \lim_{u\rightarrow 0}G_{\Delta,\ell}\sim u^{\tau/2}\sum_{m,n = 0}^\infty\frac{j^m}{\ell^{n}} \left( c^m_{n,0}K_0(2j)+ c^m_{n,1}K_1(2j) \right)
\end{equation}
with the normalisation $c^0_{0,0} = 2$ (a complete solution to the equation is outlined in Appendix \ref{app:1} and carried out in the included \texttt{Mathematica} notebook).

\begin{figure}
    \centering
    \includegraphics[width=0.4\textwidth]{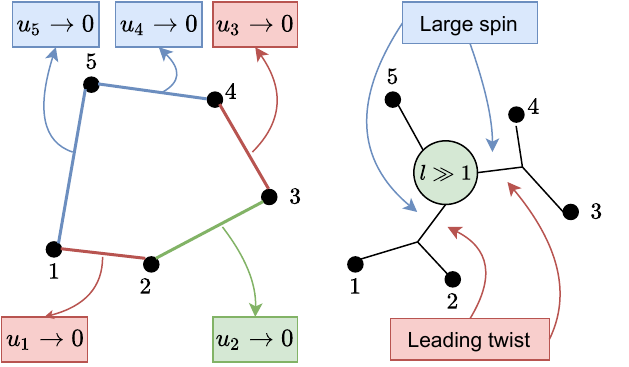}
    \caption{\label{fig:largeell}Figure borrowed from \cite{Bercini:2020msp}. Fivepoint conformal block in the lightcone limit with two large spin operators $\ell_1,\ell_2$ exchanged. The integer $l$ counts the tensor structures in the threepoint function involving these two operators and the external scalar.}
\end{figure}

The analysis of higherpoint functions can similarly reduce to a conformal block expansion---now with more cross ratios and eigenvalues---and we can play the same game \cite{Bercini:2020msp,Bercini:2021jti,Antunes:2021kmm,Bercini:2022gvs,Bercini:2024pya,Bargheer:2024hfx}. There are several different ways to choose the set of independent cross ratios, and we will make a choice that helps us take lightcone limits between pairs of points sequentially. For the fivepoint function, we choose:
\begin{equation}
    u_{1} = \frac{x^2_{12}x^2_{35}}{x^2_{13}x^2_{25}}, \quad x^2_{ij} = (x_i - x_j)^2,
\end{equation}
with $u_i = u_{i-1}|_{x_j\rightarrow x_{j+1}}$ and $x_6\equiv x_1$. The limit $u_1,u_3\rightarrow 0$ projects us into the leading twist sector of the operators exchanged in the lightcone OPE between $\mathcal{O}_1,\mathcal{O}_2$ and $\mathcal{O}_3,\mathcal{O}_4$ respectively (see Fig. \ref{fig:largeell}). This acts on the threepoint function involving two internal spinning operators and one external scalar given by \cite{Costa:2011dw}:
\begin{multline}
    \braket{\widetilde{\mathcal{O}}_1(x_2)\widetilde{\mathcal{O}}_2(x_4)\mathcal{O}_5(x_5)} = \sum_{l} C_{\ell_1,\ell_2}^{l}\times\\
    \frac{V_{2,45}^{\ell_1-l} V_{4,25}^{\ell_2-l} H_{24}^{l}}{(x_{25}^2)^{\frac{\lambda_1+\Delta-\lambda_2}{2}} (x_{45}^2)^{\frac{\lambda_2+\Delta-\lambda_1}{2}} (x_{24}^2)^{\frac{\lambda_1+\lambda_2-\Delta}{2}}}\ .
\end{multline}
Here, $\widetilde{\mathcal{O}}_1,\widetilde{\mathcal{O}}_2$ are the leading twist operators exchanged in the two channels with spins $\ell_1,\ell_2$. We also define the conformal spin variables $\lambda_i = \Delta_i + \ell_i$. Further, $V$ and $H$ are the standard tensor structures defined in \cite{Costa:2011dw} and the integer $l$ counts these structures whenever multiple spinning operators show up.

The large spin operators of interest to us, much in line with the fourpoint function analysis, dominate the double lightcone limit when we take $x_{45}^2,x_{15}^2\rightarrow 0$ (or $u_4,u_5 \rightarrow 0$). After removing the prefactor:
\begin{equation}
    \texttt{pre}_5 = \frac{2^{-\ell_1-\ell_2}\Gamma(2\ell_1+\tau_1-1)\Gamma(2\ell_2+\tau_2-1)}{\Gamma(\ell_1+\tau_1/2)^2\Gamma(\ell_2+\tau_2/2)^2}\ ,
\end{equation}
the conformal blocks in this limit simplify again and can be written as a product of Bessel functions of the form:
\begin{multline}\label{eq:fiveptblock}
    \lim_{u_4,u_5 \rightarrow 0}G_{\tau_1,\tau_2,\ell_1,\ell_2,l}(u_i)\sim u_1^{\tau_1/2} u_3^{\tau_2/2} \left(1-u_2\right)^l u_5^{\Delta_\phi/2}\times\\
    \sum_k \sum_{m_i,n_i = 0}^\infty j_1^{m_1}\ell_1^{-n_1} j_2^{m_2}\ell_2^{-n_2} (1-u_2)^k \times\\
    \left(\sum_{i_1,i_2=0}^1 c^{m_1,m_2,k}_{n_1,n_2,i_1,i_2}K_{i_1+l+\frac{\gamma_2}{2}}(2j_1) K_{i_2+l+\frac{\gamma_1}{2}}(2j_2)\right)
\end{multline}
where the sum over $k$ is cut off at a finite order depending on the large spin order $n_1+n_2$. We have defined $j_1,j_2$ similarly to $j$ in the fourpoint function case. If we further take the limit $u_2\rightarrow 0$, this cutoff changes depending on how rapidly it goes to zero as compared to $1/\ell_1,1/\ell_2$. 

In the sequel, our goal with the discussed kinematics of the large spin limit will be to extract constraints on the OPE coefficients involving multiple large spin operators using analyticity of the higher point correlators in cross ratios. It will therefore be useful to make another change of variables to make the dependence on cross ratios explicit:
\begin{equation}
    \ell_i = \frac{1}{2}\left( \frac{\sqrt{4j_i^2+u_{1-i}}}{\sqrt{u_{1-i}}} - (\tau_i-1) \right)\ .
\end{equation}
Here, it is understood that the cross ratios are cyclical in the sense that $u_0 \equiv u_5$. For the fourpoint function\footnote{Note that it is important that this change of variables is made after solving the Casimir equation since $\ell$ is fixed for each conformal block. For higher point functions, we will use the lightcone OPE to construct our blocks and can directly use $u_i, j_i$ for the expansion.}, we drop the indices and substitute $u_{1-i}\rightarrow v$.

\section{Large Spin Analysis}
The change of variables from spin $\ell$ to the rescaled conformal spin $j$ allows us to rewrite the sum over spins in the conformal block expansion into an integral over $j$. As discussed in \cite{Fitzpatrick:2012yx,Komargodski2013,Alday:2015eya}, the Bessel limit of the conformal blocks is valid to arbitrarily small $j_0$ as long as it is of order $\sqrt{v}$.

Let us now deal with the partial wave amplitudes. In the fourpoint function case, we can follow the analysis of \cite{Alday:2015eya} and define the rescaled OPE coefficients by peeling off the free theory spectrum as:
\begin{equation}
    \hat{a} = \frac{\Gamma(\ell+1+\gamma/2)\Gamma(\tau_0/2)^2}{\Gamma(\ell+\tau_0+\gamma/2-1)}\ a_{\Delta,\ell}\ .
\end{equation}
In the large spin limit, we can expect both the anomalous dimension $\gamma$ and the rescaled OPE coefficients $\hat{a}$ to have an expansion in $1/J$:
\begin{equation}
    \gamma = \sum_{n=0}^\infty \frac{v^{n/2} p_n}{j^n},\quad \hat{a} = \sum_{n=0}^\infty \frac{v^{n/2} q_n}{j^n}\ .
\end{equation}
Constraints from analyticity will then be imposed by demanding the absence of half-integer powers of the cross ratios in the conformal block expansion of the correlator. The key insight of \cite{Alday:2015eya} was that $\hat{a}(J)\left(1-\frac{\sqrt{1+4J^2}}{4J}\gamma'(J)\right)$ should only have even powers of $1/J$ in the large spin expansion. This becomes transparent once we realise that the conformal blocks in the Bessel limit only contribute even powers of $1/J$ and, due to clever choice of variables, is completely independent of anomalous dimension. The odd powers can therefore only come from the OPE coefficients and the integration measure involving derivative of the anomalous dimension.

Our goal here is to carry out the same analysis for higherpoint functions. In the fivepoint function case, the conformal block expansion for identical scalars with dimension $\Delta_\phi$ takes the form:
\begin{multline}
    \braket{\mathcal{O}_1\dots\mathcal{O}_5}=\left(\frac{1}{x_{12}^2 x_{34}^2}\right)^{\Delta_\phi} \left(\frac{x_{13}^2}{x_{15}^2 x_{35}^2}\right)^{\Delta_\phi/2}\times\\
    \sum_{\Delta_1,\Delta_2,\ell_1,\ell_2,l} a_{\Delta_1,\Delta_2,\ell_1,\ell_2}^l\ G_{\tau_1,\tau_2,\ell_1,\ell_2,l}(u_i)\ .
\end{multline}
Here, when we work in the leading twist limit, we end up retaining only the sum over $\ell_1,\ell_2,l$ and the general structure of the conformal blocks in the large spin limit is given in the previous section\footnote{In general, the odd powers of $1/J_i$ do not vanish in the conformal blocks we derive for higherpoint functions using the lightcone OPE. It is an interesting problem to find such a basis that we leave for future work.}.

After performing the change of variables to $j_1,j_2$, the sum over spins can be rewritten as an integral with appropriate measures. Note here that, in contrast with the fourpoint function case, deriving constraints requires performing the integral over $j_1,j_2$ explicitly. This is because, at any order in $1/J_i$, we get a sum of 4 terms involving different Bessel functions $K_{l+\frac{\gamma_i}{2}+m}(2j_i)$ with $m=0,1$. 

Typically, we need to do the following integral:
\begin{equation}
    \int_0^\infty dj\ j^{a} K_{b}(2j) = \frac{1}{4}\Gamma \left(\frac{a-b+1}{2} \right) \Gamma \left(\frac{a+b+1}{2}\right)
\end{equation}
which has poles when $a<b-1$. This is because we can no longer trust this integral representation of our large spin expansion, and we need to use the Casimir trick \cite{Alday:2015eya,Simmons-Duffin:2016wlq} which ends up multiplying our expansion by a factor of $j_i^2/u_{1-i}$. With the power of $j$ sufficiently positive, we can perform the integral without issues and proceed as expected.

\subsection{Constraints around tree level OPE coefficients}

We can expect that in the large spin limit the OPE coefficients have the expansion:
\begin{equation}\label{eq:ansatz}
	\hat{a}_{\ell_1,\ell_2}^l = \sum_{i_1,i_2} \hat{q}_{n_1 n_2}^\ell \left(\frac{\sqrt{u_5}}{j_1}\right)^{n_1} \left(\frac{\sqrt{u_4}}{j_2}\right)^{n_2}
\end{equation}
where we define the rescaled OPE coefficients with the following prefactor removed:
\begin{equation}
    \texttt{pre}_5' = \Gamma(l+1)^2\prod_{i=1}^{2} \frac{\Gamma\left(\frac{1}{2}(2 \ell_i-2 l-\tau_{i+1}+\tau_i+2)\right)}{\Gamma \left(\frac{1}{2} (2 \ell_i+\tau_i)\right)}
\end{equation}
defined cyclically over $i$. That is, $a_{\ell_1,\ell_2}^l=\hat{a}_{\ell_1,\ell_2}^l/\texttt{pre}_5'$. With this prefactor, the tree level OPE coefficients are simply $\hat{a}_{\ell_1,\ell_2}^l = 1$. It is easy to verify that odd powers of $1/J$ (that is $n_1+n_2$ odd) do not appear in this tree level result.

With these ingredients, we can now derive constraints on $\hat{q}_{n_1 n_2}^l$ by demanding that odd powers of the cross ratios do not appear in the expansion. The leading order constraints (valid for all values of $l$) turn out to be:
\begin{equation}\label{eq:cons1}
    \begin{gathered}
    \hat{q}_{01}^l = -l^2\ \hat{q}_{00}^{l-1} + \frac{1}{4}\left(2l+\gamma_1\right)^2\ \hat{q}_{00}^{l}\ ;\\
    \hat{q}_{10}^l = -l^2\ \hat{q}_{00}^{l-1} + \frac{1}{4}\left(2l+\gamma_2\right)^2\ \hat{q}_{00}^{l}\ .
    \end{gathered}
\end{equation}
This is one of our main results. As expected from the general principle of reciprocity, the coefficients for all odd powers in the large spin expansion are related to the even power coefficients after we have used up the constraints from lower orders. That is, the only independent coefficients are the ones with both $n_1,n_2$ even. As a consistency check, at tree level with $\gamma_i = 0$, we see that these constraints are trivial as expected. Higher order constraints can be derived programmatically.

\subsection{Trivialising Constraints at $l = 0$}

If we define a new ansatz for the OPE coefficients as:
\begin{equation}
    a_{\ell_1,\ell_2}^l = \frac{\texttt{pre}_5'}{\Gamma(l+1)^2}\ J_1^{2l+\gamma_2} J_2^{2l+\gamma_1}\ \tilde{a}_{\ell_1,\ell_2}^l\ ,
\end{equation}
with $\tilde{a}_{\ell_1,\ell_2}^l$ defined similarly to Eq. \ref{eq:ansatz}, we can again carry out the reciprocity analysis. For this case, we find the following interesting result:
\begin{equation}\label{eq:cons2}
    \tilde{q}_{n_1 n_2}^0 = 0 \quad\quad\text{when } n_1+n_2 \in 2 \mathbb{N}+1.
\end{equation}
That is, for the special case of $l=0$, we find an infinite set of constraints that are valid to all orders in the large spin expansion such that all odd powers of $1/J$ vanish identically.

For non-zero $l$, we can derive constraints order by order in $1/J$ as before. This gives the following results at leading and first subleading order:
\begin{equation}\label{eq:cons3}
    \begin{gathered}
    \tilde{q}_{01}^l = - \tilde{q}_{00}^{l-1};\qquad \tilde{q}_{10}^l = - \tilde{q}_{00}^{l-1};\\
    \tilde{q}_{11}^l = \tilde{q}_{00}^{l-2} + \frac{1}{2}\left(4l+\gamma_1+\gamma_2-2\right)\tilde{q}_{00}^{l-1}\ . 
    \end{gathered}
\end{equation}
Higher order constraints can again be derived programmatically.

Several comments are in order. Notice that, unlike the fourpoint function case, the Bessel functions appearing in the conformal block now depend on the anomalous dimensions $\gamma_i$ as well as $l$. Using the ansatz for the anomalous dimensions, we can expand the Bessel functions in powers of $1/J_i$. For example, to second order in $1/J_1$, $K_{l+\frac{\gamma_1}{2}}(2j_2)$ expands as:
\begin{gather}
    K_{l+\frac{\gamma_1}{2}}(2j_2)\sim K_{l+\frac{p_0}{2}}(2j_2)+\frac{p_1 u^{1/2}_5 K^{1,0}_{l+\frac{p_0}{2}}(2j_2)}{2j_1}\qquad\qquad\nonumber\\ + \left(\frac{p_2 u_5 K^{1,0}_{l+\frac{p_0}{2}}(2j_2)}{2j_1^2} + \frac{p_1^2 u_5 K^{2,0}_{l+\frac{p_0}{2}}(2j_2)}{8 j_1^2}\right) + \dots
\end{gather}
If we use reciprocity for the anomalous dimensions $\gamma_i$, we can see that this expansion only contains even powers of $1/J_1$. Higher order terms will therefore follow the same constraints as arising from the leading term in this expansion. These are therefore same as the constraints we derived in this work.

In the non-perturbative case, we can further expect that the leading term of the anomalous dimensions in the large spin limit is governed by the exchange of the lowest twist operator in the crossed channel and is proportional to $1/J^{\tau_{min}}$. We can therefore impose $p_0 = 0$.

We can also play the same game in perturbation theory where $p_0 \sim f(\lambda)\log(J) + g(\lambda)$ \cite{Korchemsky:1988si,Belitsky:2006en}. Here, $f$ is the cusp anomalous dimension and $g$ is the collinear anomalous dimension. For the collinear anomalous dimension, the analysis goes through as before and universal constraints can be derived that are valid to all orders in perturbation theory. In the case of the cusp anomalous dimension, because of the $\log(J)$ dependence in the Bessel functions, we can only derive constraints order by order in $\lambda$.

\section{Conclusion}
We derived constraints on the large spin expansion of OPE coefficients involving multiple spinning operators by studying the conformal block expansion of higherpoint functions. By imposing analyticity of the correlators in cross ratios, we obtained an infinite set of constraints generalising the reciprocity principle known from the fourpoint function analysis. Our constraints are valid for all tensor structures $l$. We also showed that, for the special case of $l=0$, it is possible to trivialise these constraints to all orders in $1/J_i$ by identifying a pattern in the coefficients.

The way that these constraints are derived requires knowing the expansion of the conformal blocks in the large spin limit in terms of Bessel functions to high order. In this study, we used the lighcone OPE to construct these blocks. However, this is likely not the most efficient way. As in the case of the fourpoint function, it should be possible to derive this expansion from a complete set of differential equations satisfied by the blocks. In our analysis, we were not able to do this by using Casimir equations for the five and six point cases. This is because the Casimir equations do not give good control over the polarisation vector and therefore getting a recursion relation involving the tensor structures $l_i$ is difficult. It might be more fruitful to use a more complete set of commuting operators such as the ones constructed in \cite{Buric:2020dyz,Buric:2021kgy,Buric:2021ttm,Buric:2021ywo}. The analysis of \cite{Kaviraj:2022wbw} is also particularly interesting and might help this problem. 

In a similar vein, it would also be interesting to explore analyticity of the conformal block expansion in $l$. OPE coefficients involving multiple spinning operators are known in the regime of large $l$. Relating these results to the constraints derived here for finite $l$ will be interesting. In the case of $l_i = 0$, these constraints can be trivialised to show that coefficients of odd powers in the large spin expansion of partial wave amplitudes are identically zero. Similar simplification for the constraints for general $l_i$ might warrant the construction of a more suitable basis for threepoint functions. For example, the integrability basis of \cite{Bercini:2024pya} is known to considerably simplify the perturbative data for OPE coefficients in $\mathcal{N}=4$ SYM theory. It also makes manifest the possibility of exploring analyticity in $l$. Another such basis of interest is that constructed in \cite{Costa:2023wfz}. However, these bases are not symmetric in the spins and therefore lead to an expansion depending on their ratio. This tends to make the problem of convergence of our integral representation worse. A symmetric basis that overcomes these issues would be very useful.

Another interesting line of investigation would be exploring the implication of our results on the OPE-Wilson loop duality \cite{Alday:2010ku,Alday:2010zy,Bercini:2020msp,Bercini:2021jti,Bargheer:2024hfx}. Our results are quite general and apply to any CFT in $d=4$. The only assumptions are that the external operators are identical scalars of dimension 2 which is of course not the most general case. However, it is an assumption that can be easily relaxed. In this context, it would be interesting to compare our results with \cite{Bargheer:2024hfx}. Similarly, it would also be interesting to check our predictions against loop level data for $\mathcal{N}=4$ SYM. We are also only working with the leading twist operators in our expansion. It should be possible to extend this to subleading twist operators as well. We leave these interesting problems for future work.

\acknowledgments

PA would like to thank Vasco Gon\c{c}alves for suggesting this problem, numerous discussions, and insightful comments on the draft. Centro de F\'{i}sica do Porto is partially funded by Funda\c{c}\~{a}o para a Ci\^{e}ncia e a Tecnologia (FCT) under the grant UID04650-FCUP. This work was supported by FCT grant 2024.00230.CERN.

\newpage
\appendix
\section{\label{app:1}Bessel limit of Conformal Blocks}

Conformal blocks are eigenfunctions of the Casimir operator and are labelled by the quantum numbers of the exchanged operator. Although the large spin expansion of fourpoint conformal blocks is a known result, we could not find it written out in the literature in terms of solution to the Casimir equation. It is therefore a good exercise to work it out here as it is the most efficient way we know of for calculating this expansion.

The leading twist conformal blocks $G_{\Delta,\ell}(u,v)$ satisfy the following eigenvalue equation:
\begin{gather}
    u^2 (v+1) G^{2,0}+(v-1)^2 v G^{0,2}+2 u (v-1) v G^{1,1}\qquad\qquad\nonumber\\
    \qquad\qquad+ u (v-3) G^{1,0}+ (v-1)^2 G^{0,1}
    =C_{\Delta,\ell}\ G
\end{gather}
where the exponents of $G$ tells us the order of derivative in cross ratios, and we have suppressed the arguments for brevity. The eigenvalue is given by $2 C_{\Delta,\ell} = \left((\Delta-4)\Delta+\ell^2+2 \ell\right)$. The above equation is homogeneous in $u$ and this dependence drops out if one rewrites the blocks as $G = u^{\tau/2}g(v)$. In the large spin limit, the expansion we want is in terms of the rescaled Casimir spin $j$ and cross ratio $v$. 

This is straightforward to do when we write the conformal blocks from the OPE. However, this is non-trivial from the Casimir equation perspective because $j$ itself depends on $v$. This is because the conformal blocks pick out a single family of operators labelled by $(\tau,\ell)$, and the large spin expansion mixes operators from different families. The trick is to solve the Casimir equation at fixed $(\tau,\ell)$ first by changing variables from $v$ to $j$, and then performing another change of variables from $\ell$ to $(j,v)$ in the final result. The equation we need to solve is written in the attached \texttt{Mathematica} notebook.

The solution can be found by substituting the ansatz of Eq. \ref{eq:fourptansatz}. The coefficients $c^m_{n,0},c^m_{n,1}$ can be found order by order in $1/\ell$ and $j$ in terms of recursion relations. At each order in $1/\ell$, the initial conditions we need to provide are for $c^0_{n,0},c^1_{n,1}$. The reason is that at each level $n$, the recursion relations are such that these terms enter as $m^2 c^m_{n,0}$ and $(m-1)^2 c^m_{n,1}$ respectively. This means that even though we have $2mn$ coefficients to determine, we only get $2(m-1)n$ equations from the recursion relations. We therefore need to suppliment the system with $2n$ boundary conditions.

From normalisation of our blocks, we know the leading coefficient $c^0_{0,0} = 2$. This is the solution to part of the Casimir equation that is homogeneous in $j$. We further assume that $c^0_{n,0}=0$ for $n > 0$, which is justified by the blocks derived using the OPE.

Finding the relevant boundary conditions for $c^1_{n,1}$ is more tricky. Our solution is saved by realising that these conditions can be generated equivalently by imposing conditions on $c^m_{n,0}$. That is, we can impose the condition for cutting off the summation in $m$ beyond the highest power of $j$ that appears for a given order in $\ell$. From inspecting the blocks coming from the OPE, we know that the summation in $m$ goes up to $ 2\left(\left\lfloor \frac{n}{2}\right\rfloor +\left\lfloor \frac{n}{4}\right\rfloor \right)$. This exact cutoff is however not important as long as we impose a cutoff that is strictly larger. For example, $c^6_{2,0} = 0$ gives $c^3_{2,1} = -4/3$, which in turn gives $c^1_{2,1} = -2/3$. We can also impose $c^4_{2,0} = 0$ to get the same result. With this, we can solve for all coefficients in the expansion. We verify our calculation by matching with the known expansion from the OPE up to order $1/\ell^{14}$.

\bibliographystyle{JHEP}
\bibliography{conformald.bib}

\end{document}